\documentclass{aa}
\usepackage{txfonts}
\usepackage{graphicx}
\usepackage{natbib}
\bibpunct{(}{)}{;}{a}{}{,}
\newcommand{\ltsim}{\raise 2pt \hbox {$<$} \kern-1.1em \lower 4pt \hbox {$\sim$}}
\newcommand{\gtsim}{\raise 2pt \hbox {$>$} \kern-1.1em \lower 4pt \hbox {$\sim$}}

\begin{document}

\title{B2 1144+35B, a giant low power radio galaxy \\
       with superluminal motion.}

\subtitle{Orientation and evidence for recurrent activity}

\author{G. Giovannini\inst{1,2} \and 
M. Giroletti \inst{2} \and
G.B. Taylor\inst{3}} 

\institute{Dipartimento di Astronomia, Universita' di Bologna,
via Ranzani 1, 40127 Bologna, Italy
\and
Istituto di Radioastronomia - INAF, via Gobetti 101, 40129 
Bologna, Italy \and
Department of Physics and Astronomy, University of New Mexico,
Albuquerque, NM 87131, USA}

\abstract
{}
{The goal of this work is a detailed study of the nearby, low power 
radio galaxy B2 1144+35B.}
{For this purpose, we performed new Very Long Baseline Array (VLBA) and 
Very Large Array (VLA) observations.}  
{This source has several properties (bright, nearby, large range of
spatial structures, visible counterjet, etc.) that make it an
excellent astrophysical laboratory for the study of the evolution of
radio jets.  Here we report the detection of motion in the counterjet
at 0.23 $\pm$ 0.07 c, which allows us to estimate the orientation of
the parsec-scale jet at 33$^\circ \pm 7^\circ$ from the line of sight, 
with an intrinsic velocity of
(0.94$^{+0.06}_{-0.11}$)c.  We also report on a brightening of the core at high
frequencies which we suggest could be the result of a new component
emerging from the core.  High dynamic range VLBA observations at 5 GHz
reveal, for the first time, extended emission connecting the core with
the bright complex of emission that dominates the flux density of the
parsec scale structure at frequencies below 20 GHz.  The evolution of
this bright complex is considered and its slow decline in flux density
is interpreted as the result of an interaction with the interstellar
medium of the host galaxy.}
{}


\maketitle

\section{Introduction}

The low power radio galaxy \object{B2 1144+35B} (hereafter referred to as
1144+35) is identified with a faint (m$_{pg}$ = 15.7) Zwicky galaxy (ZW186.48)
in a medium-compact galaxy cluster at a redshift of 0.0630.

In an optical study of bright, flat-spectrum radio sources, \citet{ma96}
classify 1144+35 as a BL Lac candidate even though its spectrum shows H$\alpha$
and [NII] emission lines \citep{co75,mo92}.  From a comparison between the
measured line equivalent width and the line contrast, \citet{ma96} suggest that
1144+35 could be a diluted BL Lac. 1144+35 has also been included in an imaging
and spectroscopic survey of low and intermediate power radio galaxies
\citep{eb89}.  A CCD residual image shows a very definite arc of dust in the
nuclear region of the host galaxy.

From the radio point of view, 1144+35 has a peculiar structure as discussed in
detail by \citet{g99}.  It is characterized by a large scale FR-I radio
structure and \citet{mac98} classifies it among the giant radio galaxies. The
total angular size is $\sim$ 13\arcmin\ corresponding to a projected linear
size of $\sim$0.9 Mpc. The total radio power at 1.4 GHz is 6.3 $\times 10^{24}$
W Hz$^{-1}$.  Despite its large size, the kiloparsec scale radio structure is
core dominated with a short, bright two-sided jet.  The arcsecond core exhibits
long-term flux density variability. Its flux density showed a large increase
from 1974 to 1980 followed by a smooth increase until 1992. From 1992 until
1999 the arcsecond core flux density has been decreasing \citep{g99}.

The first VLBI observations of this source were carried out by \citet{g90}, and
successive VLA, MERLIN and VLBI data \citep{g99} reveal complex structure over
a broad range of physical scales (1 pc -- 1 Mpc).  The bright arcsecond scale
core is resolved at milliarcsecond resolution into a nuclear source, a main jet
with an apparent superluminal velocity, and a faint counter-jet (cj).

We present here new multi-frequency VLBA observations (\S2) to study the source
evolution, proper motion, flux density variability, and radio spectral index
(\S3). Moreover we present flux density measurements of the arcsecond core in
order to compare its variability with the parsec scale source
evolution. Results are discussed in \S4, and conclusions are given in \S5.

Assuming H$_0$ = 70 km sec$^{-1}$ Mpc$^{-1}$, $\Omega_m$ = 0.3 and 
$\Omega_\Lambda$ = 0.7, the luminosity distance (D$_l$) for 1144+35 is
282.7 Mpc, the angular distance (D$_a$) is 250.2 Mpc with a conversion factor 
= 1.21 kpc/\arcsec.

\begin{table*}
\caption{Arcsecond core flux densities}
\label{tab1}
\centering
\begin{tabular}{c c c c c c c}
\hline\hline
Epoch & Flux(mJy) & Flux(mJy & Flux(mJy) & Flux(mJy) & Flux(mJy) & Flux(mJy) \\
mm-yy & 1.4 GHz & 5.0 GHz & 8.3 GHz & 15 GHz & 22 GHz & 43 GHz \\
\hline
03-02 & 399     &  330    &  263    &   226  &  200   &  146    \\
06-06 & 338     &  243    &  222    &   204  &  193   &  153    \\
\hline
\end{tabular}
\end{table*}

\section{Observations and Data Reduction}

\subsection{VLBI Data}

New observations have been obtained with the VLBA of the NRAO\footnote{The
National Radio Astronomy Observatory is operated by Associated Universities,
Inc., under cooperative agreement with the National Science Foundation} on
February 17, 2002 at 5 and 8.4 GHz and on September 22, 2005 at 1.5, 5, and 8.4
GHz.  Observations were carried out, switching often between different
frequencies in order to obtain good and uniform ($u,v$) coverage.  The data
were correlated in Socorro with the VLBA correlator.  Amplitude calibration was
performed using observations of the system temperature and antenna gains in the
standard way.  Data were globally fringe-fitted and calibrated in AIPS.  We
made several iterations of phase self-calibration, followed by a final phase
and gain self-calibration when necessary to produce the final image.

Observations at 8.4 GHz (2005 epoch) were performed in a phase-referencing mode
in order to obtain a good core position useful for future studies of absolute
motion. We used J1130+3815 as primary calibrator, and J1130+3031, J1214+3309 as
secondary calibrators.

\subsection{VLA Data}

We obtained a $\sim$ 2 hour observation with the VLA (NRAO) in the A (March
2002) and A/B configuration (June 2006) to investigate the flux density
variability of the arcsecond scale core.  The data have been calibrated in the
standard way using the NRAO AIPS package and imaged using the task IMAGR. In
these images due to the lack of short spacings and the short integration time,
only the unresolved arcsecond core is visible and therefore we make use of
these data only to discuss the variability of the arcsecond core.  Observations
were performed at 6 different frequencies and the core flux density for each is
reported in Table 1. The core flux density was derived by fitting a Gaussian to
the nuclear source in the image plane.

\section{Results}

\subsection{Parsec scale morphology, variability, and spectral indices}

The new observations confirm the general structure reported in \citet{g99}.  We
label the different components as shown in Fig. 1, in accordance with
\citet{g99}.  The parsec scale structure at 5 and 8 GHz (Figs. 1 and 2) is
resolved in a nuclear source (C) with a short jet (D) and cj structure (E). The
core position from the phase reference data is: R.A. = 11 47 22.12902 Dec = +35
01 07.5350 (J2000).  At about 25 mas east of the core we detect an extended jet
like structure (A and B components) which dominates the flux density in the
VLBI images.  This jet structure is extended, clearly limb-brightened, and
connected to the core by a low-brightness emission. The opening angle is $\sim$
10$^\circ$, in agreement with the value found by \citet{g99}.  The shape of
this extended feature is similar in different epoch observations.  In
particular the high-resolution images at 8.4 GHz confirm that component A1 is
resolved into an extended complex structure aligned with components C, D, and
E.  Furthermore this complex structure curves in the direction of component A,
suggesting a possible jet recollimation, most evident in the last epoch.
Moreover a fairly compact component B2, not present in 1997 images, has emerged
just west of component B1. The flux density of compact sub-structures is not
constant.

\begin{figure*}
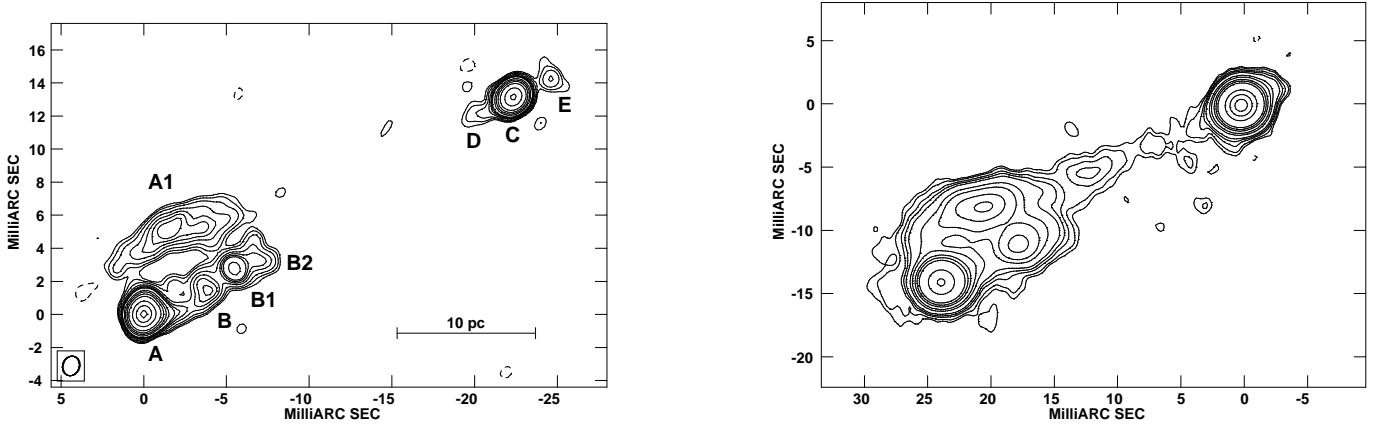

\centering
\includegraphics[width=0.45\linewidth]{8058f1a.ps}
\hfill
\includegraphics[width=0.45\linewidth]{8058f1b.ps}
\caption{VLBI images of 1144+35 at 8.4 GHz. 
Left: 2002 epoch; the HPBW is 1.2 $\times$ 1.0 mas in PA -20$^\circ$; 
the noise level
is 0.1 mJy/beam. Contours are: $-$0.5 0.5 0.7 1 1.5 2 2.5 3 5 7 10 15 30 50
70 mJy/beam. Letters indicate different components as discussed in the text.
Right: 2005 epoch; the HPBW is 2 $\times$ 2 mas; 
the noise level is 0.05 mJy/beam. 
Contours are drawn at:
$-$0.3 0.2 0.3 0.5 0.7 1 2 3 5 7 10 30 50 70 mJy/beam.}
\label{f1ab.ps}
\end{figure*}

At 5 GHz and in the low-resolution 8.4 GHz images, for the first time, 
the connection between the two-sided core region 
and the distant jet
structure is readily visible confirming the peculiar alignment 
between components
A1, D, C, and E while components A, B, B1, and B2 are slightly off axis. 
A low-brightness emission in front of component A
is present, probably connected to component A1.

At lower frequency (1.4 GHz, Figure 3) no low brightness emission is visible at
larger distances from the core. The faint ($\sim$ 1 mJy/beam) extended feature
seen in 1995 observations at $\sim$ 90 mas from the core \citep{g99}, aligned
with the A+B structure, if real, could have been missed because of lower
brightness due to expansion.  The extension just in front at the compact A
component is confirmed.  The jet is transversally resolved suggesting the
presence of a low brightness steep spectrum emission surrounding the brighest
jet regions visible in high resolution maps at 5 and 8.4 GHz.

In Table 2 we report the measured flux density at different epochs and
frequencies for each component.  Components are named according to Figure 1. JT
is the total flux density of the main jet (A, A1, B, B1, B2 components). It can
be slightly higher than the sum of different components because of low surface
flux density in between the single substructures. The flux density variability
is shown in Figure 4 for the main components. We show only 8.4 GHz data, since
at this frequency better statistics are available, and we have the high angular
scale images necessary to separate the various components.

\begin{table}
\caption{Flux densities at various epochs of VLBI components}
\label{tab2}
\centering
\begin{tabular}{c c c c c}
\hline\hline
Component & Epoch & 8.4 GHz & 5 GHz & 1.4 GHz \\
          &  yr   & mJy     & mJy   & mJy     \\
\hline
 A & 1995.22 & 255    &  -    &   -       \\
 A & 1995.90 & 224    &  -    &   -       \\
 A & 1997.74 & 184    &  -    &   -       \\
 A & 2002.13 & 103    &  147  &   -       \\
 A & 2005.72 &  63    &   68  & 130       \\
 B & 1995.22 & 28.7   &  -    &   -       \\
 B & 1995.90 & 37.0   &  -    &   -       \\
 B & 1997.74 & 21.8   &  -    &   -       \\
 B & 2002.13 & 11.9   &  -    &   -       \\
 B & 2005.72 & 2.9    &  6.2  &   -       \\
 C & 1995.22 & 59.7   &  -    &   -       \\
 C & 1995.90 & 62.6   &  -    &   -       \\
 C & 1997.74 & 64.9   &  -    &   -       \\
 C & 2002.13 & 53.7   &  50.1 &   -       \\
 C & 2005.72 & 74.0   &  43.4 & 24.6      \\
A1 & 1995.22 & 41.5   &  -    &   -       \\
A1 & 1995.90 & 34.9   &  -    &   -       \\
A1 & 1997.74 & 43.7   &  -    &   -       \\
A1 & 2002.13 & 27.5   &  42.3 &   -       \\
A1 & 2005.72 & 29.3   &  32.5 &   -       \\
B1 & 2002.13 & 9.6    &   -   &   -       \\
B1 & 2005.72 & 2.0    &  6.7  &   -       \\
B2 & 2002.13 & 5.6    &  -    &   -       \\
B2 & 2005.72 & 7.5    &  6.2  &   -       \\ 
JT & 2002.13 & 164    & 232   &   -       \\
JT & 2005.72 & 115    & 125   & 269       \\
\hline
\end{tabular}

\end{table}

We derived the spectrum of individual components using data at the 2005.72
epoch when three different frequencies were available.  We confirm the results
published in \citet{g99}: i) component C is the only self-absorbed component
with the self-absorption frequency \gtsim ~~8.4 GHz ($\alpha^{8.4}_{1.4}$ =
-0.61 $\pm$0.03, assuming S($\nu$) $\propto$ $\nu^{-\alpha}$); ii) the spectral
shape of component A, despite the large variation in flux density, did not
change: $\alpha^{5.0}_{1.4}$ = 0.51 $\pm$0.02 with a high frequency flattening
($\alpha^{8.4}_{5.0}$ = 0.15 $\pm$0.04) in agreement with the compact structure
of this component; iii) other components show a moderately steep spectrum. The
average spectrum of the whole strong jet component is $\alpha^{8.4}_{1.4}$
$\sim$ 0.5.

\begin{figure}
\resizebox{\hsize}{!}{\includegraphics{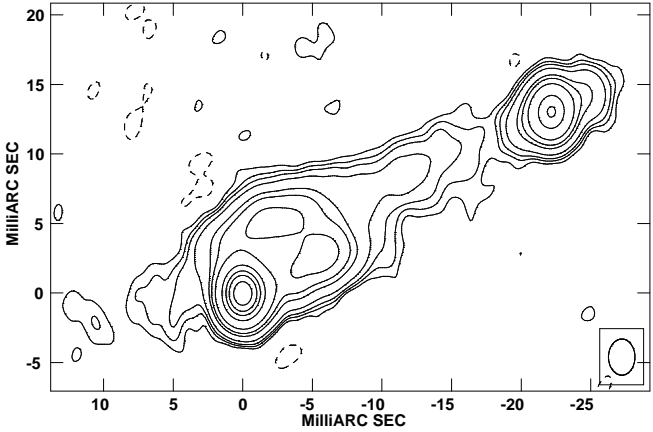}}
\caption{VLBI image of 1144+35 at 5 GHz (2005 epoch).
The HPBW is 2.6 $\times$ 1.9 mas in PA 0$^\circ$;
the noise level is 0.06 mJy/beam. Contours are drawn at: $-$0.15 0.15 0.3 0.5 
1 3 5 10 30 
50 70 100 mJy/beam.}
\label{f2.ps}
\end{figure}

\begin{figure}
\resizebox{\hsize}{!}{\includegraphics{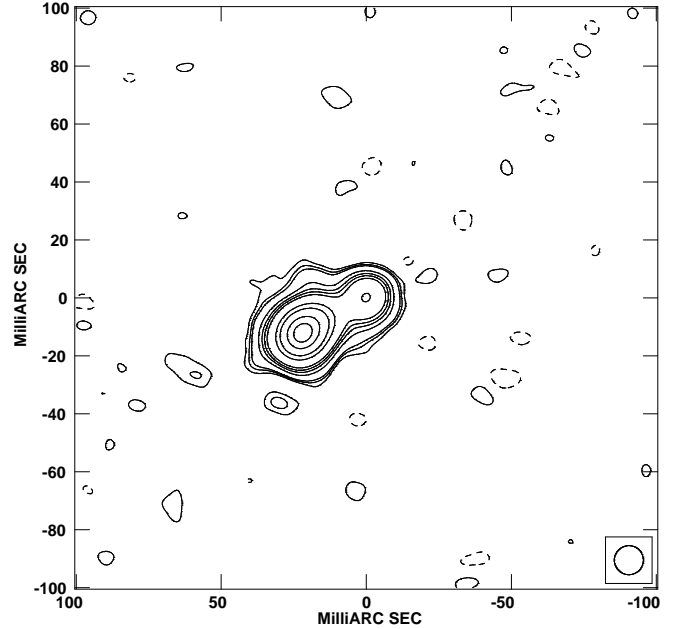}}
\caption{VLBI image of 1144+35 at 1.5 GHz (2005 epoch). The HPBW is 
10 $\times$ 10 mas;
the noise level is 0.14 mJy/beam. Contours are drawn at: $-$0.5 0.4 0.7 1 3 5 
7 10 30 
50 100 150 mJy/beam.}
\label{f3.ps}
\end{figure}

\begin{figure}
\resizebox{\hsize}{!}{\includegraphics{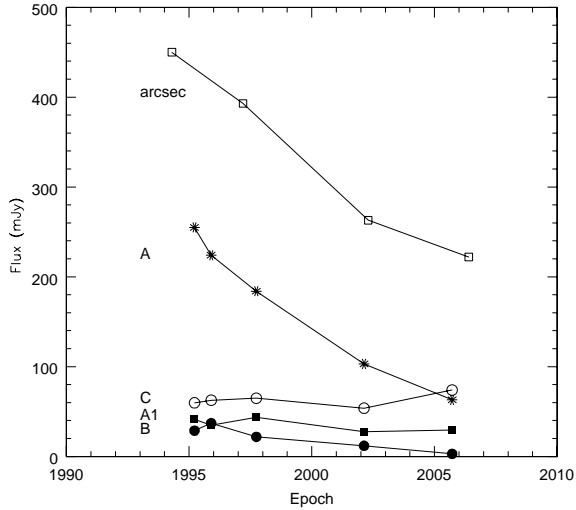}}
\caption{Flux density measures at 8.4 GHz at different epochs of the
arcsecond core and of parsec scale components named according to Figure 1.
Connecting 
lines are drawn for display purpose. Components B1 and B2 are omitted for
clarity and because of their low flux density. Flux density errors are 2\% or 
0.1 mJy for fainter components.}
\label{f4.ps}
\end{figure}

\begin{figure}
\resizebox{\hsize}{!}{\includegraphics{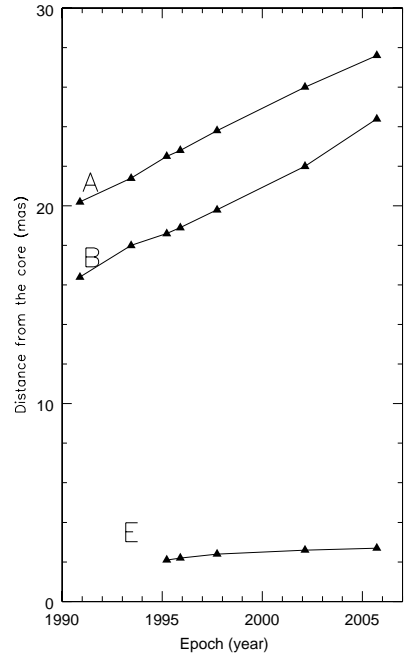}}
\caption{Distance of A, B, and E components from the parsec scale core (C) at 
different epochs. Connecting lines are for display only.}
\label{f5.ps}
\end{figure}

\begin{figure}
\resizebox{\hsize}{!}{\includegraphics{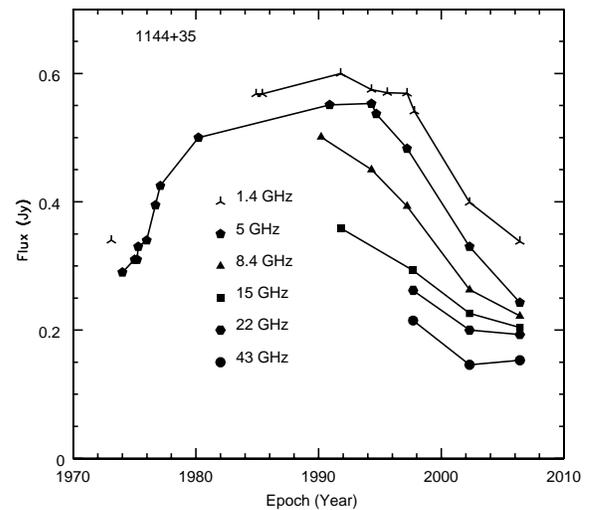}}
\caption{Core flux density at arcsecond resolution, at different epochs and 
frequencies. Lines connecting points are for
display only.}
\label{f6.ps}
\end{figure}

Based on its spectral properties and compactness we confirm component C as the
center of activity for 1144+35B.

Using 5 and 8.4 GHz data obtained at the same epoch we produced images with the
same ($u,v$) range, cellsize and angular resolution. We used these data to
produce a spectral index image. The spectral index image confirms the
properties of single components as discussed above, moreover it shows that the
short jet and cj components (D and E) have a steep spectrum ($\alpha \sim$ 1.0)
and that also the low brightness structure connecting D to A1 is very steep
($\alpha$ in the range 1.1 - 1.6).

\subsection{Proper Motion}

We used the observations made at different epochs to measure the apparent
proper motion of components A, B, and E with respect to the core component C.
We note that new observations confirm the result discussed in \citet{g99} that
the relative positions of main jet sub-structures appear constant in time. Also
the radio emission centroid of component A1 in different epoch images is always
in between the A and B components, suggesting a similar apparent velocity for
all the bright jet structure. The position of the short jet D is poorly defined
being smooth with no substructure, therefore it is not possible to measure a
reliable proper motion for it.

In Table 3 and Figure 5 we show the distance of components A, B, and E from C
at different epochs. For component E we use only high resolution 8.4 GHz data.
By considering all observations to date we measure for the first time a proper
motion of the cj (E) structure in agreement with the previous upper limit:
$\beta_{acj}$ = 0.23 $\pm$ 0.07, while we confirm $\beta_{aj}$ = 1.92 $\pm$
0.05 for components A and B (note the different cosmology used here with
respect to previous papers).

\begin{table}
\caption{Distances from the core at different epochs}
\label{tab3}
\centering
\begin{tabular}{c c c c c}
\hline\hline
Epoch & E   & A   & B   & Frequency \\
year  & mas & mas & mas & GHz       \\
\hline
1990.89 & -   &  20.2 & 16.4 & 5.0 \\
1993.44 & -   &  21.4 & 18.0 & 5.0 \\
1995.22 & 2.1 &  22.5 & 18.6 & 8.4 \\
1995.90 & 2.2 &  22.8 & 18.9 & 8.4 \\
1997.74 & 2.4 &  23.8 & 19.8 & 8.4 \\
2002.13 & 2.6 &  26.0 & 22.0 & 8.4 \\
2005.72 & 2.7 &  27.6 & 24.4 & 8.4 \\
\hline
\end{tabular}
\end{table}

\subsection{The arcsecond core}

The arcsecond core of 1144+35 is the dominant feature of the radio emission
from this galaxy and has long been known to be variable \citep{ek83}.  The
arcsecond core flux densities available in the literature have been presented
in \citet{g99}.  We add here new recent measurements (Table 1). In Figure 6 we
present the plot of the arcsecond core flux density at different epochs and
frequencies.

We derive the core spectral index by comparing observations at the same time or
very near in time.  We note that from 1990 until 2002 we have a flat spectrum
on the lower frequency range ($\alpha_{1.4}^{5.0}$) which however is steepening
in time (0.10 in 1990, 0.22 in 2002). In the higher-frequency range (5 to 15 or
43 GHz) the spectral index is constant in time and slightly steeper:
$\alpha_{5.0}^{43}$ = 0.38. In the 2006 epoch we have a significant change with
a flattening in the higher frequency range: ($\alpha_{1.4}^{5.0}$ = 0.26 and
$\alpha_{5.0}^{43}$ = 0.21).  Uncertainties in the above spectral index values
are $\sim$ 0.03.

\section{Discussion}

\subsection{Jet Orientation and Velocity}

From the measured apparent proper motion of the parsec scale jet and cj,
assuming intrinsic symmetry, we can derive $\beta cos\theta$ for this parsec
scale jet independently of the Hubble constant following \citet{mi94}:

\vskip 0.5truecm

\centerline {$(\beta_{\rm aj} - \beta_{\rm acj}$)/$(\beta_{\rm aj} + 
\beta_{\rm acj})$ = $\beta cos\theta$}

\vskip 0.5truecm
We find: 

\centerline{$\beta cos\theta$ = 0.79 $\pm$ 0.07}
\vskip 0.5truecm

Since we know the source D$_a$ in the present cosmology we can also derive the
intrinsic jet orientation \citep{mi94}:

\vskip 0.5truecm

\centerline{tan$\theta$ = (2D$_a$ $\mu_j \mu_{cj})/(c(\mu_j - \mu_{cj}))$}
\vskip 0.5truecm

\noindent
where $\mu_j$ and $\mu_{cj}$ are respectively the jet and cj angular
velocity from proper motion measures.
 
We find $\theta$ = 33$^\circ$ $\pm$ 7$^\circ$ and therefore $\beta$ = 0.94
$^{+0.06}_{-0.11}$ is the intrinsic value of the jet velocity in c
units.\footnote{Note that we did not assume an error for the Angular Distance.}
These values agree with the measured jet - cj arm ratio ($\sim$ 10) which
implies \citep{g98,ta97}:

\vskip 0.5truecm

\centerline{$\beta cos\theta$ $\sim$ 0.8.}
\vskip 0.5truecm

Therefore the pattern jet velocity from proper motion is in agreement with the
jet bulk velocity derived from the jet - cj arm ratio.  The Lorentz factor
$\gamma$ is 2.9 and the Doppler factor $\delta$ is $\sim$ 1.6.

\subsection{Parsec scale jet morphology}

In light of previous results and in agreement with the discussion given in
\citet{g99}, the limb-brightened jet morphology can be explained by a
transversal velocity structure. The observed brightness of the external jet
regions is amplified (the Doppler factor is $\delta$ $\sim$ 1.6). If we assume
that the inner jet spine is moving at higher velocity (e.g. ~~ $\beta \sim$
0.998), the spine Doppler factor will be $\delta$ $\sim$ 0.39, therefore the
observed brightness of the inner spine is dimmed and the jet will appear
limb-brightened. A similar structure was found in the BL Lac object Markarian
501 \citep[see][]{gir04}.

In this scenario the gap between the core and the A+B complex is explained by
the low value ($<$ 1) of the Doppler factor before the velocity decrease of the
outer sheath.  The jet sub-structures are regions with a strong interaction
between the jet and the surrounding medium, with a jet velocity decrease and a
higher observed brightness. The large asymmetry between A1 and A+B+B1+B2 side
could reflect a different interaction between the jet and an inhomogeneous ISM
\citep[see][for a more detailed discussion]{g99}.

We note that a jet velocity structure is readily detected only in sources in a
narrow range of orientation with respect to the line sight: in sources near the
plane of sky only the external shear is visible, while in sources at a small
angle with respect to the line of sight, the fast spine is highly boosted and
dominates the radio images.

\subsection{Flux density variability}

The arcsecond core flux density of 1144+35 shows a slow, but well defined
variability (Figure 6).  The core flux density increases from 1974 to $\sim$
1994 and decreases from 1994 to 2002 at all frequencies.  From a comparison of
the flux density variability of arcsecond and VLBI components (see Figure 4 and
Tables 1 and 2), it is clear that the flux density decrease is not due to the
VLBI core but to changes in the compact jet component A.  The reason of the
strong flux density decreases of component A is not clear:

\begin {itemize}

\item
The decrease cannot be due to adiabatic expansion since in all epochs this
component is slightly resolved in images at 8.4 GHz and we did not measure any
increase of its size;

\item
We exclude a flux density decrease because of radiation losses related to the
aging of relativistic particles, because of the too short time scale and the
constant spectrum at different epochs (moreover the flux density decrease is
present also at 1.4 GHz);

\item
The decrease cannot be due to a change of the Doppler factor because of a
different orientation angle with respect to the line of sight, or a jet
velocity decrease since the proper motion of the whole bright jet structure is
regular and constant in time.

\end{itemize}

It could be due to a change of the external medium: component A being the most
distant component of the jet substructure could be interacting with a dense
cloud and local reaceleration processes have increased its brightness. In this
scenario, around 1995 the A component emerged from a cloud and its brightness
started to decrease. A dynamic interaction of the whole jet structure (A + B
components) is supported also by the small changes in its morphology such as
the presence of a new sub-component (B2) with respect to previous images.  This
is a speculative, if real we should see a change in the proper motion velocity:
slower before (during the interaction with the cloud) and faster after the
interaction. Unfortunately we started to monitor the parsec scale structure
when the flux density was already decreasing (after the cloud interaction) so
that we cannot yet put this model to the test.

We note that in the 2002 - 2006 period the arcsecond core flux density decrease
stopped in the high frequency range (we have a flux density increase at 43 GHz
and a constant value at 22 GHz). This change in the flux density variability is
not due to the A component (see Figure 4), but to an increase in the VLBI core
flux density, which became in 2006 the strongest component on the VLBI scale.
A possible interpretation of this result is that the core C began a new active
phase with the emission of a new component not yet visible in our images
because of its small size and short distance from the core. This component is
still self absorbed at frequencies lower than 22 GHz and we find indication of
its existence only at 22 and mostly at 43 GHz. Therefore only the arcsecond
core flux density measured at high frequency is at present affected by this new
component.  Assuming a self-absorption frequency above $>$ 8.4 GHz, and
equipartition conditions, the size of this new component is $<$ 0.03 mas.  If
this component is moving at the same expanding velocity of the jet, we should
start to resolve it in VLBI images at 8.4 GHz in \gtsim ~~2 years.
  
\subsection{Source evolution}

The large scale structure shows a clear discontinuity between the extended
relaxed lobes and the high brightness arcsecond core and jets.  These
observational data suggest different phases in the life of this radio source:
the extended structure is a relic emission with an age in the range 5 to 9
$\times$ 10$^7$ yrs as estimated in \citet{g99}, while the emission on the
arcsecond-scale has a shorter dynamical age: assuming an average velocity of
\gtsim ~~0.02 c for the arcsecond structure we derive an age of \ltsim ~~1.0
$\times 10^7$ yr, taking into account projection effects.

We note also that the parsec scale jet emission does not decrease smoothly in
our VLBI maps. The main parsec scale jet appears to stop at the end of the A
component at $\sim$ 30 mas from the core. Assuming a constant velocity (see
Figure 5) the main VLBI jet structure was emitted from the core a few decades
ago, circa 1950.  Moreover we see from VLBI images the main component A is
followed by a complex bright structure with many subcomponents suggesting that
the source activity related to the emission of the A component lasted about 10
- 20 years with an almost continuous ejection.  Presumbably source activity and
jet production was also present in epochs before 1950, since there are
evidences from MERLIN data that a low level flux density emission is present at
larger distances from the core, in the same position angle.  Moreover the short
arcsecond-scale jet shows a high brightness well defined structure. We can
suppose that multiple activity phases happened in the last 10$^7$ yrs.  In this
scenario we can add the evidence discussed in the previous sub-section that a
new component was emitted in $\sim$ 2002 and should be visible at 8.4 GHz in a
couple of years, giving us the opportunity to study its evolution in connection
with the arcsecond-scale flux density variability.

Finally we note that the new data do not add any information to understanding
the reason for the asymmetry between the A1 and A+B jet structures.  This
remains a puzzle.

\section{Conclusions}

New observations discussed here confirm the complex parsec-scale structure of
the source 1144+35. This parsec-scale structure is well-defined and we can
easily follow it in different epochs. However some changes are evident, in
particular the flux density decrease of the A component and the presence of a
new fairly compact component B2 in the main jet structure, suggesting a dynamic
interaction with the surrounding medium and possible local particle
reacceleration by turbulence. Such interactions may trigger star formation
\citep{cro06}, and influence the evolution of the host galaxy.  Feedback
mechanisms between the growth of the central black hole and the galaxy
evolution have been shown to be important in large elliptical galaxies in
clusters \citep{all06}, and could be important in lower luminosity systems as
well.

A few peculiarities have not yet been understood: 1) high sensitivity images
reveal the connection between the core C and the bright jet substructure
confirming the peculiar alignment between components A1, D, C, and E while
other bright components (A, B, B1, B2) are slightly off axis. 2) The low
brightness emission in front of component A is in contrast with our favored
scenario in which the bright spot A is the working surface of the jet impinging
on the surrounding medium. 3) The evident asymmetry between components A1 and
the A, B, B1, and B2 complex, possibly due to a different interaction with the
ISM, implies different properties in the ISM on the parsec-scale. 4)
Investigation into the low brightness extended emission surrounding the bright
jet components will require higher sensitivity data before it can be properly
discussed.

The study of the flux density variability in this source is important to
understand the small-scale structures. The recent increase in the core
(component C) flux density at high frequency is interpreted as evidence of a
new jet component expelled from the core a few years ago and now moving along
the jet.  Further along the jet, the large flux density decrease of the
arcsecond core from 1992 to 2002 is due mainly to the jet component A, but the
physical mechanism is still unknown.

A spectral index study confirms the core identification with the only
self-absorbed component and shows that compact sub-components have a flatter
spectrum with respect to extended low brightness regions.

The proper motion of jet components is regular and constant in time.  The
detection of a cj proper motion allows us to derive the intrinsic jet velocity
and orientation with respect to the line of sight. These results are in
agreement with the estimate of the bulk jet velocity derived from the jet to cj
arm ratio.

The radio morphology of this source shows clear discontinuities at different
linear scales suggesting a continuous activity but with high and low level
periods. The connection between this continuous, but not constant, radio
activity of the central AGN with the evolution of the parent galaxy is not yet
understood.

The radio galaxy 1144+35 is an interesting {\it astrophysical laboratory} to
study parsec scale jet properties and evolution and their connection with the
kiloparsec scale structure, because of its low redshift, complex but well
studied structure, constant but slow variability in flux density and
morphology, a well detected jet and counter-jet proper motion, and a large
amount of available data.

If we compare this source with other low-power radio galaxies, we find that the
measured jet velocity and orientation agrees with unified model predictions and
with general parameters derived from statistical studies
\citep[e.g. see][]{g01}. 1144+35 is one of the few sources where the jet
velocity from proper motion is in agreement with the jet velocity from the jet
- cj arm ratio.  However some properties of this source are quite peculiar and
not yet understood as discussed before.  In particular the level of radio
activity in this source is clearly not constant in time. Hints of recurrent
activity have been found in other low-power radio galaxies, but sufficiently
detailed studies for a comparison to 1144+35 are lacking.  We can conclude
that, while some general properties of low-power radio sources are at present
well-established (e.g. existence of relativistic jets on the parsec scale, jet
velocity decrease with core distance, connection between pc and kpc scale
jets), properties of restarted radio sources are still poorly known and require
deep and detailed studies.

\begin{acknowledgements}

We thank Dr. Luigina Feretti for useful discussions and suggestions, and an
anonymous Referee for useful and important comments.  This work was partially
supported by the Italian Ministry for University and Research (MIUR) and by the
National Institute for Astrophysics (INAF).

\end{acknowledgements}

\end{document}